\documentstyle[12pt]{article}
\def\baselinestretch{1.5}
\def\vecsq{\stackrel{{}_{\rightarrow {\scriptstyle 2}}}}

\newcommand{\be}{\begin{equation}}
\newcommand{\ee}{\end{equation}}
\newcommand{\ba}{\begin{array}}
\newcommand{\ea}{\end{array}}
\begin{document}
%\psnormal
%\small
\begin{titlepage}
\vspace*{-1cm}
\hfill{hep-ph/9212305}\\
\phantom{bla}
\hfill{IEM-FT-64/92}\\
\phantom{bla}
\hfill{December 1992}
\vskip 0.5cm
\normalsize
\begin{center}
{\Large\bf Upper Bounds on the Lightest Higgs Boson Mass in General
Supersymmetric Standard Models $^{\star}$}
\end{center}
\vskip .5cm
\begin{center}
{\bf J. R. Espinosa}$^{\dagger}$ and {\bf M. Quir\'os}$^{\ddagger}$\\
Instituto de Estructura de la Materia\\
Serrano 123, 28006 Madrid, Spain
\end{center}

\begin{abstract}
\noindent
In a general supersymmetric standard model there is an upper bound
$m_h$ on the tree level mass of the $CP=+1$ lightest Higgs boson
which depends on the electroweak scale, $\tan \beta$ and the gauge
and Yukawa couplings of the theory. When radiative corrections are
included, the allowed region in the $(m_h,m_t)$ plane depends on
the scale $\Lambda$, below which the theory remains perturbative,
and the supersymmetry  breaking scale $\Lambda_s$, that we fix to
$1\ TeV$. In the minimal model with $\Lambda=10^{16}\ GeV$:
$m_h<130\ GeV$ and $m_t<185\ GeV$. In non-minimal models with an
arbitrary number of gauge singlets and $\Lambda=10^{16}\ GeV$:
$m_h<145\ GeV$ and $m_t<185\ GeV$. We also consider supersymmetric
standard models with arbitrary Higgs sectors. For models whose couplings
saturate the scale $\Lambda=10^{16}\ GeV$ we find $m_h<155\ GeV$ and
$m_t<190\ GeV$. As one pushes the saturation scale $\Lambda$ down to
$\Lambda_s$, the bounds on $m_h$  and $m_t$ increase. For instance,
in models with $\Lambda=10\ TeV$, the upper bounds for $m_h$ and
$m_t$ go to $415\ GeV$ and $385\ GeV$, respectively.
\end{abstract}
\vspace{3cm}
\footnoterule
\def\baselinestretch{.5}
\large
{\footnotesize
$^{\star}$ Work partly supported by
CICYT under contract AEN90-0139.\newline
$^{\dagger}$ Supported by a grant of Comunidad de
Madrid, Spain.\newline
$^{\ddagger}$ e-mail:quiros@roca.csic.es.}
\end{titlepage}

%\vskip2truecm
%%%%%%%%%%%%%%%%%%%%%%%%%%%%%%%%%%%%%%%%%%%%%%%%%%%%%%%%%%%%%%%%%%%%%%
\def\baselinestretch{1.5}
\large
\normalsize
%\nsect{}
The most outstanding challenge for present (Tevatron, LEP) and future
(LEP-200, NLC, LHC, SSC) colliders is the discovery of the Higgs boson
\cite{1}, which might confirm the standard model as the final theory
of the electroweak interactions. However, though the standard model
(SM) is in excellent agreement with all precision measurements at
present energies \cite{1,2}, extensions thereof are not excluded at
higher scales. The most appealing of these extensions, which provides
a technical solution to the hierarchy puzzle, is the supersymmetric
standard model \cite{3}.

Supersymmetric models have well constrained Higgs sectors \cite{4}
which can provide crucial tests of them. In particular, the most
constraining feature of the minimal supersymmetric standard model
(MSSM) is the existence of an absolute upper bound on the tree-level
mass of the $CP=+1$ lightest Higgs boson
\be
m_h\leq m_Z\mid \cos 2\beta\mid,
\ee
where $\tan\beta\equiv v_2/v_1$, $v_i\equiv\langle H_i^o\rangle$. Therefore a
negative result on the Higgs search would seem to exclude
phenomenological supersymmetry at all making its search
at future accelerators unnecessary. However relation $(1)$ is
spoiled by two effects: i) Radiative corrections, and ii) The
enlargement of the Higgs sector in non-minimal supersymmetric
standard models (NMSSMs).
Only the simultaneous consideration of both effects can provide
reliable bounds in general supersymmetric models.

Radiative corrections have been computed in the MSSM by different
groups using different methods: standard diagrammatic
techniques \cite{5}, the one-loop effective potential \cite{6} and the
renormalization group (RG) approach including one \cite{7} and
two \cite{8} loop corrections. All approaches provide remarkably
coincident results. The latter is reliable provided that
\be
\Lambda_s^2/m_W^2\gg 1,
\ee
where $\Lambda_s$ is the scale of supersymmetry breaking, since it
amounts to a resummation of all leading logarithms in the effective
potential. On the other hand this method is universal because we
are assuming that the standard model holds below $\Lambda_s$ and the
supersymmetric theory beyond $\Lambda_s$. Under the condition $(2)$
supersymmetric particles decouple from the low-energy theory and the
RG procedure is expected to provide a good enough description of the
radiative contribution to the lightest Higgs boson mass in a general
supersymmetric standard model. Our definition of {\it lightest Higgs
boson} is the $CP=+1$ bosonic state
whose mass is not controlled by $\Lambda_s$ in the sense that it has
a finite limit when $\Lambda_s\rightarrow\infty$. In that limit
supersymmetry decouples and the latter state becomes the SM Higgs boson.
We will include radiative corrections for $\Lambda_s=1\ TeV$ using
the RG approach \cite{8}. The radiative squared mass $\Delta m^2_r$ is
$\beta$-dependent and has to be added to the tree-level mass.

The tree-level bound $(1)$ does not hold in NMSSMs. The case of the
MSSM plus a singlet with coupling $\lambda$ to $H_1\cdot H_2$ was
first studied in \cite{9,10} where a tree-level bound was found as
\be
m_h^2\leq \left(\cos^2 2\beta +\frac{2\lambda^2\cos^2
\theta_W}{g^2}\sin^2 2\beta\right)m_Z^2,
\ee
which is $\Lambda_s$-independent. From $(3)$ we see that the bound on
$m_h$ is linked to the bound on $\lambda$ if we require the theory to
remain perturbative between $\Lambda_s$ and $\Lambda$. For
$\Lambda=\Lambda_{GUT}$, the unification scale of gauge coupling
constants, the bound $(3)$ was studied in \cite{11,11'} and \cite{12},
where radiative corrections were properly included. The case of one
extra singlet has been recently reconsidered in \cite{13}, where the
dependence of $m_h$ on $\Lambda$ was studied; \cite{14}, where
radiative corrections where considered in the effective potential
approach; and \cite{15}, where comparison with the corresponding
non-supersymmetric scenario was established. All of these results
agree, when they overlap, with our previous calculation \cite{12}
within less than $5\%$. More general models, {\it e.g.} the MSSM plus any
number of singlets or three $SU(2)$ triplets (whose vacuum
expectation values (VEVs) can respect the custodial symmetry at
tree-level), were presented by ourselves in \cite{11,12} and,
more recently, also considered in \cite{16}.

In this paper we will present upper bounds on the lightest Higgs
boson mass in a general class of models:
supersymmetric standard models with an arbitrary Higgs sector. We will
assume:
\newpage
\begin{itemize}
\item Two doublets $H_1^{(1)}$, $H_2^{(1)}$, with hypercharges
$Y=\pm 1/2$, coupled to quarks and leptons in the superpotential
\be
f_m=h_t  Q\cdot H_2^{(1)} U^c + h_b  Q\cdot H_1^{(1)} D^c+
h_{\tau} L\cdot H_1^{(1)} E^c,
\ee
plus an arbitrary number of extra pairs $H_1^{(j)}$,
$H_2^{(j)}$, $j=2,...,d+1$, decoupled from quarks and leptons in
order to avoid dangerous flavor changing neutral currents \cite{17}.
\item Gauge singlets $S^{(\sigma)}$, $\sigma=1,...,n_s$.
\item $SU(2)$ triplets $\Sigma^{(a)}$, $a=1,...,t_o$, with $Y=0$.
\item $SU(2)$ triplets $\Psi^{(i)}_1$, $\Psi^{(i)}_2$,
$i=1,...,t_1$, with $Y=\pm 1$.
\end{itemize}

Notice that the above extra Higgses are the only ones that can
provide renormalizable couplings to $H_1^{(1)}\cdot H_2^{(1)}$,
$H_1^{(1)} H_1^{(1)}$ and $H_2^{(1)} H_2^{(1)}$ in the
superpotential. Other (more exotic) Higgs representations will only
contribute to the $\beta$-functions of the gauge couplings and give
lower values of the upper bound $m_h$. Since we are only interested
in absolute upper bounds we can disregard them. Therefore the above
class of extra Higgs fields defines the most general Higgs sector in
supersymmetric standard models as far as the issue of putting upper
bounds on their lightest Higgs boson mass is concerned.

The most general renormalizable superpotential for the above Higgses
can be written as $f_h+g$ where:
\be
\ba{c}
f_h= \lambda_{1}^{ij\sigma} H_1^{(i)}\cdot H_2^{(j)} S^{(\sigma)}
+  \lambda_{2}^{ija} H_1^{(i)}\cdot \Sigma^{(a)} H_2^{(j)}
 +  \chi_{1}^{jkb} H_1^{(j)}\cdot \Psi_1^{(b)} H_1^{(k)} \\  +
 \chi_{2}^{jkb} H_2^{(j)}\cdot \Psi_2^{(b)} H_2^{(k)}, \vspace{.2cm} \\
\ea
\ee
\be
\ba{c}
g=  \lambda_{ajk} tr(\Sigma^{(a)}\Psi_1^{(j)}\Psi_2^{(k)}) +
\frac{1}{6}\kappa_{abc} tr(\Sigma^{(a)}\Sigma^{(b)}\Sigma^{(c)})
+\frac{1}{6}\kappa_{\sigma\mu\nu} S^{(\sigma)}S^{(\mu)}S^{(\nu)}.
\ea
\ee
Summation over repeated indices is understood and
traces are taken over matrix indices in the field
decomposition:
\be
\ba{c}
H_1=\left(\begin{array}{c}
H_1^o\\
H_1^-
\end{array}\right),\hspace{1cm}
H_2=\left(\begin{array}{c}
H_2^+\\
H_2^o
\end{array}\right)\vspace{.4cm},
\ea
\ee
\be
\ba{c}
\Sigma=\left(\begin{array}{cc}
\xi^o/\sqrt{2}&-\xi_2^+\\
\xi_1^-&-\xi^o/\sqrt{2}
\end{array}\right),\vspace{.4cm}
\ea
\ee
\be
\ba{c}
\Psi_1=\left(\begin{array}{cc}
\psi_1^+/\sqrt{2}&-\psi_1^{++}\\
\psi_1^o&-\psi_1^+/\sqrt{2}
\end{array}\right),\vspace{.4cm}
\ea
\ee
\be
\ba{c}
\Psi_2=\left(\begin{array}{cc}
\psi_2^-/\sqrt{2}&-\psi_2^o\\
\psi_2^{--}&-\psi_2^-/\sqrt{2}
\end{array}\right)\vspace{.3cm}.
\ea
\ee
where we have dropped the multiplicity indices.

By making a unitary transformation in $j$ space to the Higgs doublets
$H_1^{(j)}, H_2^{(j)}$  we can assume, without loss of
generality, that only $H_1^{(1)}$ and $H_2^{(1)}$ take a non-zero
VEV \cite{18}. This requires the cancellation of various Yukawa couplings in
eq. (5). In particular we find the condition
\be
\lambda_{1}^{1j\sigma}=\lambda_{2}^{1ja}=\chi_{1}^{1jb}=
\chi_{2}^{1jb}=0 \hspace{.3cm} (j\neq 1) ,
\ee
that will be assumed from here on.

After imposing condition (11) we can write $f_h=f+...$, where the
ellipsis involves only fields which do not acquire VEV, {\it i.e.}
$\langle f_h\rangle=\langle f\rangle$, as
\be
f=\vec{\lambda}_{1}\cdot\vec{S} H_1^o H_2^o-
\frac{\vec{\lambda}_{2}}{\sqrt{2}}\cdot\vec{\xi}^o H_1^o H_2^o+
\vec{\chi}_{1}\cdot\vec{\psi}_1^o H_1^o H_1^o+
\vec{\chi}_{2}\cdot\vec{\psi}_2^o H_2^o H_2^o.
\ee
Where we use the notation
\be
\ba{cl}
(\vec{\lambda}_{1})^{\sigma}=\lambda_{1}^{11\sigma},&\vspace{.4cm}\\
(\vec{\lambda}_{2})^{a}=\lambda_{2}^{11a},&\vspace{.4cm}\\
(\vec{\chi}_{i})^{b}=\chi_{i}^{11b},&\vspace{.4cm}\\
H_i^o=H_i^{(1)o} &(i=1,2) .
\ea
\ee

Using the superpotential (12), and the fact that the smallest eigenvalue
of a real, symmetric $n\times n$ matrix is smaller than the smallest
eigenvalue of the upper left $2\times 2$ submatrix, one can easily prove
that the lightest Higgs boson mass has an upper bound given by
\cite{11}
\be
m_h^2/v^2\leq\frac{1}{2}(g^2+g'^2)\cos^2 2\beta +
(\vecsq{\lambda_{1}}
+\frac{1}{2}\vecsq{\lambda_{2}})\sin^2 2
\beta+ \vecsq{\chi_1}\cos^4\beta+ \vecsq{\chi_2}\sin^4\beta,
\ee
where $v^2\equiv v_1^2+v_2^2$ and $g,g'$ are the $SU(2)\times U(1)_Y$
gauge couplings. In particular the bound (3) is recovered when
$\vec{\lambda}_{2}=\vec{\chi}_{1}=\vec{\chi}_{2}=0$ and the bound (1)
in the MSSM when also $\vec{\lambda}_{1}=0$.

As it has been repeatedly noticed \cite{11,16} the bound (14) is
independent of the soft-supersymmetry breaking
parameters, or any supersymmetric mass terms. It
is only controlled by $v$ and dimensionless parameters
($g,g',\tan\beta$, and Yukawa couplings). Since the former is
fixed by the electroweak scale, the latter will determine the bound
(14). In particular the upper bound on the right hand side of (14)
comes from the requirement that the supersymmetric theory remains
perturbative below $\Lambda$. To guarantee this condition we need to
solve the renormalization group equations (RGEs) of all gauge and
Yukawa couplings of the theory.

In the general supersymmetric theory defined by the superpotential
(4-6) with the condition (11), the relevant RGEs to the bound (14) can be
written, for the Yukawa couplings as:
\be
\ba{cl}
4\pi^2\displaystyle{\frac{d}{dt}}\vecsq{\lambda_1}&=
\left( -\frac{3}{2}g^2 -\frac{1}{2}g'^2 +\frac{3}{2}(h_t^2+h_b^2)
+\frac{3}{2}\vecsq{\lambda_2}
+3(\vecsq{\chi_1}+\vecsq{\chi_2})+
2\vecsq{\lambda_1}\right)\vecsq{\lambda_1}\vspace{.3cm}\\
&+\displaystyle{\sum_{k,l\neq 1}}
(\lambda_{\sigma}^{11}\lambda_{\sigma}^{kl})^2+
\frac{1}{4}\displaystyle{\sum_{\mu,\nu}}
(\lambda_{\sigma}^{11}\kappa_{\sigma \mu\nu})^2,
\vspace{.4cm}\\
\ea
\ee
\be
\ba{cl}
4\pi^2\displaystyle{\frac{d}{dt}}\vecsq{\lambda_2}&=
\left( -\frac{7}{2}g^2 -\frac{1}{2}g'^2 +\frac{3}{2}(h_t^2+h_b^2)
+2\vecsq{\lambda_2}
+3(\vecsq{\chi_1}+\vecsq{\chi_2})+
\vecsq{\lambda_1}\right)\vecsq{\lambda_2}\vspace{.3cm}\\
&+\displaystyle{\frac{1}{2}\sum_{k,l\neq 1}}
(\lambda_{a}^{11}\lambda_{a}^{kl})^2
+\frac{1}{2}\displaystyle{\sum_{k,l}}
(\lambda_{a}^{11}\lambda_{akl})^2
+\frac{1}{4}\displaystyle{\sum_{k,l}}
(\lambda_{a}^{11}\kappa_{akl})^2,
\vspace{.4cm}\\
\ea
\ee
\be
\ba{cl}
4\pi^2\displaystyle{\frac{d}{dt}}\vecsq{\chi_1}&=
\left( -\frac{7}{2}g^2 -\frac{3}{2}g'^2 + 3 h_b^2
+\frac{3}{2}\vecsq{\lambda_2}
+7 \vecsq{\chi_1}+
\vecsq{\lambda_1}\right)\vecsq{\chi_1}\vspace{.3cm}\\
&+\displaystyle{\frac{1}{2}\sum_{a,r}}
\left( \chi_{1}^{11i}\lambda_{air}\right)^2
+\displaystyle{\sum_{m,r\neq 1}}
\left(\chi_{1}^{11i}\chi_{1}^{mri}\right)^2,
\vspace{.4cm}\\
\ea
\ee
\[
\ba{cl}
4\pi^2\displaystyle{\frac{d}{dt}}\vecsq{\chi_2}&=
\left( -\frac{7}{2}g^2 -\frac{3}{2}g'^2 + 3 h_t^2
+\frac{3}{2}\vecsq{\lambda_2}
+7 \vecsq{\chi_2}+
\vecsq{\lambda_1}\right)\vecsq{\chi_2},\vspace{.3cm}\\
\ea
\]
\be
\ba{cl}
&+\displaystyle{\frac{1}{2}\sum_{a,r}}
\left(\chi_{2}^{11i}\lambda_{ari}\right)^2
+\displaystyle{\sum_{m,r\neq 1}}
\left(\chi_{2}^{11i}\chi_{2}^{mri}\right)^2,
\vspace{.4cm}\\
\ea
\ee
\be
\ba{cl}
8\pi^2\displaystyle{\frac{d h_t}{dt}}&=
\left( -\frac{3}{2}g^2 -\frac{13}{18}g'^2 -\frac{8}{3}g_s^2
+3h_t^2+ \frac{1}{2}h_b^2
+\frac{3}{4}\vecsq{\lambda_2}
+3\vecsq{\chi_2}+
\frac{1}{2}\vecsq{\lambda_1}\right)h_t\vspace{.3cm}\\
&+\left(\displaystyle{\frac{3}{4}\sum_{k\neq 1}}\lambda_{2}^{1ka}
\lambda_{2}^{1ka}+3\displaystyle{\sum_{k\neq 1}}\chi_{2}^{k1i}
\chi_{2}^{k1i}+ \displaystyle{\frac{1}{2}\sum_{k\neq
1}}\lambda_{1}^{k1\sigma}
\lambda_{1}^{k1\sigma}\right)h_t,\vspace{.4cm}
\ea
\ee
\be
\ba{cl}
8\pi^2\displaystyle{\frac{d h_b}{dt}}&=
\left( -\frac{3}{2}g^2 -\frac{7}{18}g'^2 -\frac{8}{3}g_s^2
+3h_b^2+ \frac{1}{2}h_t^2
+\frac{3}{4}\vecsq{\lambda_2}
+3\vecsq{\chi_1}+
\frac{1}{2}\vecsq{\lambda_1}\right)h_b\vspace{.3cm}\\
&+\left(\displaystyle{\frac{3}{4}\sum_{k\neq 1}}\lambda_{2}^{k1a}
\lambda_{2}^{k1a}+3\displaystyle{\sum_{k\neq 1}}\chi_{1}^{1ki}
\chi_{1}^{1ki}+ \displaystyle{\frac{1}{2}\sum_{k\neq
1}}\lambda_{1}^{1k\sigma}
\lambda_{1}^{1k\sigma}\right)h_b,\vspace{.4cm}
\ea
\ee
and for the gauge couplings as:
\be
\ba{cl}
16\pi^2\displaystyle{\frac{d g}{dt}}&=(1+2t_0+4t_1+d)g^3,
\ea
\ee
\be
\ba{cl}
16\pi^2\displaystyle{\frac{d g'}{dt}}&=(11+6t_1+d)g'^3,
\ea
\ee
\be
\ba{cl}
16\pi^2\displaystyle{\frac{d g_s}{dt}}&=-3g_s^3,
\ea
\ee
where $g_s$ is the $SU(3)$ gauge coupling
\footnote{For simplicity we did not present the
explicit RGEs for the couplings $\lambda_1^{ij\sigma},\
\lambda_2^{ija},\ \chi_1^{ijb},\ \chi_2^{ijb}$,
$\lambda_{ajk}$, $\kappa_{abc}$ and $\kappa_{\sigma \mu \nu}$.}.
Notice that condition (11) is stable under the RGEs.

The $\tau$-Yukawa coupling, $h_{\tau}$, will be neglected as compared
to $h_b$, since $h_{\tau}/h_b=m_{\tau}/m_b$. The bottom Yukawa
coupling can be important for $\tan\beta\gg 1$ and will be kept along
with the top Yukawa coupling. They are given by
\be
\ba{c}
h_t=\displaystyle{\frac{g}{\sqrt{2}}\frac{m_t}{m_W}}
(1+\cot^2\beta)^{1/2},\vspace{.3cm}\\
h_b=\displaystyle{\frac{g}{\sqrt{2}}\frac{m_b}{m_W}}(1+\tan^2\beta)^{1/2},
\ea
\ee
and fixed by the boundary conditions: $m_t(2m_t)=m_t$ and
$m_b(2m_b)=5\ GeV$. For the gauge couplings we will take the
boundary conditions:
\be
\ba{ccc}
\alpha_{EM}(M_Z)=\displaystyle{\frac{1}{127.9}},&\sin^2\theta_W(M_Z)=0.23,&
\alpha_s(M_Z)=0.12\ .
\ea
\ee

The Yukawa couplings involved in (14) will be let to acquire any
perturbative value maximizing the bound (14). In general the bound
(14) is maximized whenever some of the involved couplings {\it saturate} the
scale $\Lambda$. A particular coupling $\lambda$ (gauge or Yukawa) is
said to {\it saturate} a scale $\Lambda$ if
\be
\lambda^2(Q^2)/4\pi\leq 1,
\ee
for $Q^2\leq \Lambda^2$,
and the equality in (26) holds for $Q^2= \Lambda^2$.

Once we have the set of RGEs (15-23) we can systematically analyze
the case of {\mbox different} supersymmetric standard models
characterized by
different Higgs sectors. The different cases are parametrized by the
number of Higgs representations $(n_s,d,t_0,t_1)$. For simplicity we
will assume $t_0=t_1\equiv t$ such that the custodial symmetry (and so
the tree level value $\rho\equiv 1$) can be respected by means of
a very simple relation between the VEVs of $\Sigma, \Psi_1$ and $\Psi_2$.
In this way the gauge couplings will be parametrized by the single
parameter
\be
N=6t+d,
\ee
and the bound will depend on $n_s$ and $N$. In all cases the bound will
be a function of $\Lambda$, and so it is important to discuss the criteria
to fix it.
\begin{itemize}
\item In theories with gauge coupling unification at the scale
$\Lambda_{GUT}$, the natural choice is fixing
$\Lambda=\Lambda_{GUT}$. The case satisfying this condition is the
MSSM plus an arbitrary number of gauge singlets, {\it i.e.} $N=0$, $n_s$
arbitrary. Of course putting $n_s=0$ we recover
the MSSM.
\item In theories without gauge coupling unification, $N\neq 0$, we
will assign to any scale $\Lambda$ the value of $N$ such that the
gauge coupling $g$ saturates it \footnote{We can see from
(21-23) and conditions (25-27) that every scale $\Lambda$
is always saturated by $g$: this means that $g$ is the first gauge
coupling to go non-perturbative.}.
Once we have fixed $N$, the bound (14) is maximized for $n_s\neq 0$,
and so we will assume there are gauge singlets. The actual value
of the bound does not depend on the particular value of $n_s$,
provided that $n_s>0$.
\end{itemize}

The key observation to maximize the bound (14) is to notice that
$\vecsq{\lambda_i}$ and $\vecsq{\chi_i}\ (i=1,2)$ are maximized by
the values
\be
\ba{cc}
\lambda_{1}^{ij\sigma}=\lambda_{2}^{ija}=\chi_{1}^{ijb}=
\chi_{2}^{ijb}=0& (i,j\neq 1), \vspace{.3cm}\\
\lambda_{ajk}=\kappa_{abc}=\kappa_{\sigma\mu\nu}=0,&
\ea
\ee
since they contribute to the renormalization of $\vecsq{\lambda_i}$
and $\vecsq{\chi_i}$ $(i=1,2)$ with positive definite terms. Eq. (28) is
stable under the RGEs, so we can impose it in order to obtain upper
bounds on $m_h$. It might happen in specific models that we would need to
departure from condition (28) to encompass the low energy
experimental bounds. For instance, this could be the case when some of the
couplings involved in (28) are necessary to explicitly break a
global symmetry; in that case, to avoid a massless axion, we should
put them to non-zero values. However, since we are dealing only
with absolute upper bounds we will use condition (28) in the
rest of this paper.

We have analyzed different cases characterized by different values of
$\Lambda$. In {\mbox Fig$.\, 1$} we plot $m_h$ as a function of $m_t$ for the
model $N=0$, $n_s>0$ (i.e. the MSSM enlarged with Higgs singlets) and
different values of $\tan\beta$. In Fig$.\, 2$ we plot the case
saturating the scale $\Lambda=10^{16}\ GeV$, which corresponds to
$N=5$. The model with $t=1, d=0$ $(N=6)$, which saturates the scale
$\Lambda=10^{14}\ GeV$, is shown in Fig$.\, 3$. We see that the value of
$m_h$ increases as we let the saturation scale $\Lambda$ to go down
to $\Lambda_s$.
In particular we show in Fig$.\, 4$ the plot corresponding to
$\Lambda=10^{10}\ GeV$ $(N=10)$. Finally we show in Fig$.\, 5$ the
allowed region in the $(m_h,m_t)$ plane for the MSSM (long-dashed); the
model $N=0,n_s>0$ with $\Lambda=\Lambda_{GUT}$ (short-dashed), and
different models saturating different scales, from $\Lambda=10^{16}\ GeV$
to $\Lambda=10\ TeV$ (solid).

In conclusion, we have computed upper bounds on the mass of the
lightest Higgs boson in a general class of supersymmetric standard
models, characterized by arbitrary Higgs sectors.
For a given scale $\Lambda$ the corresponding upper bound is
saturated by the model whose gauge couplings become non-perturbative
at that scale. We have taken the scale of supersymmetry breaking
$\Lambda_s=1\ TeV$ and assumed a supersymmetric theory
between $\Lambda$ and $\Lambda_s$ and the standard model
below $\Lambda_s$. In detailed studies of particular models,
threshold effects of supersymmetric particles should be taken into
account, as well as physical conditions on the Yukawa couplings that
we are not considering ({\it e.g.} unbroken color and/or electric
charge). These effects could slightly change our absolute upper
bounds. Radiative corrections are taken into account using the
(universal) renormalization group method. Diagrammatic
techniques should give similar results provided that
$\Lambda_s^2\gg m_W^2$. (This has been explicitly checked in the
simplest non-minimal model with a singlet.) For lower values of
$\Lambda_s$ the detailed calculation should be done for every
particular model. We did not analyze the dependence of the upper
bounds on $\Lambda_s$ but we expect it not to be too dramatic provided
$\Lambda_s$ is kept inside its phenomenological range. For all the
above reasons our results should be taken as rough estimates to guide
the eye in supersymmetric standard models. They are strongly dependent
on the scale $\Lambda$. In fact for $\Lambda$ from $10^{16}\ GeV$ to
$10^4\ GeV$ the upper bounds on $m_h$ range from 130 to $415\ GeV$,
and the bounds on $m_t$ from 185 to $385\ GeV$. For $\Lambda =\Lambda_s$ the
bound coincides with that in the non-supersymmetric standard model \cite{19}.

We have not considered in this paper other possible generalizations
of the MSSM that could, in turn, modify the upper bounds on the
lightest Higgs boson mass. One of them is the introduction of extra
colored multiplets, as {\it e.g.} extra (fractions of) generations.
They would provide large radiative corrections, through the Yukawa
couplings of the additional up quarks, to the Higgs boson mass for
the case of a not-so-heavy top quark \cite{20}. Another possible
generalization
is the presence of an extra gauge group $G$, with couplings $g_a$ and
generators $T^a$ \cite{21}, which would amount to a correction to the
tree-level bound (14) as
\be
\Delta m_h^2= 2g_a^2v^2 \sum_a\left( T_1^a \cos^2\beta - T_2^a
\sin^2\beta\right)^2
\ee
where $T^a_i\equiv \langle H_i^{(1)}T^aH_i^{(1)}\rangle/v_i^2$. In that case
the Yukawa couplings in (4) and (5) should respect the gauge symmetry
$G$ and all the bounds could change dramatically.

\newpage
\def\baselinestretch{1.2}
\large
\normalsize
\section*{Figure Captions}
\begin{description}
\item[Fig. 1] Upper
bounds on the lightest scalar Higgs boson in NMSSM with singlets and
different values of $\tan \beta$ (solid).
\item[Fig. 2] The same as in Fig.1 but for a model saturating
$\Lambda =10^{16}\ GeV$.
\item[Fig. 3] The same as in Fig.1 but for a model saturating
$\Lambda =10^{14}\ GeV$.
\item[Fig. 4] The same as in Fig.1 but for a model saturating
$\Lambda =10^{10}\ GeV$.
\item[Fig. 5] Allowed regions in the $(m_h,m_t)$ plane for different
supersymmetric standard models. The solid curves correspond to models
saturating the scales $\Lambda =10^{16},10^{14},10^{10},10^8,10^6$
and $10^4\ GeV$.
\end{description}
\end{document}